\documentclass[prb,reprint,superscriptaddress,longbibliography]{revtex4-1}
\usepackage{graphicx}
\usepackage{amssymb}
\usepackage{amsmath}
\usepackage[product-units = power,separate-uncertainty=true]{siunitx}
\usepackage{amsfonts}
\usepackage{bbold}
\usepackage[version=3]{mhchem}
\usepackage[citecolor=blue,colorlinks=true]{hyperref}
\usepackage[dvipsnames]{xcolor}
\usepackage{nicefrac}

\DeclareSIUnit\BohrMagneton{$\mu_{\textrm{B}}$}
\DeclareSIUnit\formulaunit{f.u.}
\DeclareSIUnit\atomicunit{a.u.}
\DeclareSIUnit\arbunit{arb.unit}
\DeclareSIUnit\torr{Torr}
\DeclareSIUnit\counts{Counts}
\DeclareSIUnit\rlu{r.l.u.}
\DeclareSIUnit\unitcell{u.c.}

\usepackage{pgfplots}
\usepackage{tikz}
\usepgfplotslibrary{external}
\usepgfplotslibrary{groupplots}
\usetikzlibrary{plotmarks}
\usetikzlibrary{positioning}
\usepackage{pgfplotstable}
\usetikzlibrary{calc}
\tikzsetexternalprefix{figures/}
\pgfplotsset{select coords between index/.style 2 args={
    x filter/.code={
        \ifnum\coordindex<#1\fi
        \ifnum\coordindex>#2\fi
    }
}}

\usetikzlibrary{arrows.meta}
\usetikzlibrary{decorations.markings}
\tikzset{test/.style n args={3}{
    postaction={
    decorate,
    decoration={
    markings,
    mark=between positions 0 and \pgfdecoratedpathlength step 0.5pt with {
    \pgfmathsetmacro\myval{multiply(
        divide(
        \pgfkeysvalueof{/pgf/decoration/mark info/distance from start}, \pgfdecoratedpathlength
        ),
        100
    )};
    \pgfsetfillcolor{#3!\myval!#2};
    \pgfpathcircle{\pgfpointorigin}{#1};
    \pgfusepath{fill};}
}}}}

\usetikzlibrary{backgrounds}
\definecolor{graphicbackground}{rgb}{1,1,1}
\pgfkeys{/tikz/.cd,
  background color/.initial=graphicbackground,
  background color/.get=\backcol,
  background color/.store in=\backcol,
}
\tikzset{background rectangle/.style={
    fill=\backcol,
  },
  use background/.style={
    show background rectangle
  }
}

\definecolor{c1}{rgb}{0.2,0.8,0.3} 
\definecolor{c2}{rgb}{0.8,0.3,0.2} 
\definecolor{c3}{rgb}{0.85,0.33,0.1} 
\definecolor{c4}{rgb}{0.7,0.1,0.5} 
\definecolor{c5}{rgb}{0.3,0.2,0.8} 

\definecolor{teal}{rgb}{0.0,0.5,0.5} 
\definecolor{navy}{rgb}{0.0,0.0,0.5} 
\definecolor{purple}{rgb}{0.5,0.0,0.5} 
\definecolor{darkorange}{rgb}{1.0,0.55,0.0} 
\definecolor{goldenrod}{rgb}{0.85,0.64,0.125} 
\definecolor{celticsgold}{rgb}{0.54,0.43,0.3} 
\definecolor{hotpink}{rgb}{1.0,0.41,0.7} 
\definecolor{mediumvioletred}{rgb}{0.78,0.08,0.52} 
\definecolor{turquoise}{rgb}{0.25,0.875,0.81} 
\definecolor{darkturquoise}{rgb}{0.0,0.8,0.82} 
\definecolor{forestgreen}{rgb}{0.13,0.54,0.13} 
\definecolor{seagreen}{rgb}{0.19,0.54,0.34} 
\definecolor{celticsgreen}{rgb}{0,0.48,0.2} 
\definecolor{steelblue}{rgb}{0.27,0.51,0.70} 
\definecolor{dodgerblue}{rgb}{0.12,0.56,1} 
\definecolor{mediumblue}{rgb}{0.0,0.0,0.8} 
\definecolor{warriorsblue}{rgb}{0,0.412,0.71} 
\definecolor{royalblue}{rgb}{0,0.324,0.73} 
\definecolor{hornetspurple}{rgb}{0.11,0.066,0.375} 
\definecolor{yellowgreen}{rgb}{0.68,1.0,0.18} 
\definecolor{mediumseagreen}{rgb}{0.23,0.70,0.44} 
\definecolor{wheat}{rgb}{0.96,0.87,0.70} 

\usepackage{hyperref}
\hypersetup{
    bookmarks=true,         
    bookmarksopen=true,
    unicode=true,          
    pdftoolbar=true,        
    pdfmenubar=true,        
    pdffitwindow=false,     
    pdfstartview={FitH},    
    pdfborder={0 0 0},      
    pdftitle={DyBCO charge order using resonant x-ray scattering},    
    pdfauthor={Davide Betto et al.},     
    pdfsubject={DyBCO charge order with RXS},   
    pdfcreator={Davide Betto},   
    pdfkeywords={RXS, Charge Order, Charge Density Waves, Cuprates}, 
    pdfnewwindow=true,      
    colorlinks=true,       
    linkcolor=red,          
    citecolor=mediumseagreen,        
    filecolor=magenta,      
    urlcolor=steelblue,           
    linktocpage=true,       
    hyperindex=true,        
    pdfpagelabels=true,
    plainpages=false,
}

\begin{document}

\title{Imprint of charge and oxygen orders on Dy ions in \ce{DyBa2Cu3O_{6+x}} thin films probed by resonant x-ray scattering}
\author{Davide Betto}
\email[Email address: ]{d.betto@fkf.mpg.de}
\affiliation{Max Planck Institute for Solid State Research, Heisenbergstra{\ss}e 1, 70569 Stuttgart, Germany}
\author{Martin Bluschke}
\affiliation{Max Planck Institute for Solid State Research, Heisenbergstra{\ss}e 1, 70569 Stuttgart, Germany}
\affiliation{Helmholtz-Zentrum Berlin f\"{u}r Materialien und Energie, Albert-Einstein-Stra{\ss}e 15, 12489 Berlin, Germany}
\author{Daniel Putzky}
\affiliation{Max Planck Institute for Solid State Research, Heisenbergstra{\ss}e 1, 70569 Stuttgart, Germany}
\author{Enrico Schierle}
\affiliation{Helmholtz-Zentrum Berlin f\"{u}r Materialien und Energie, Albert-Einstein-Stra{\ss}e 15, 12489 Berlin, Germany}
\author{Andrea Amorese}
\affiliation{Max Planck Institute for Chemical Physics of Solids, N\"othnitzer Stra{\ss}e 40, 01187 Dresden, Germany}
\author{Katrin F\"{u}rsich}
\affiliation{Max Planck Institute for Solid State Research, Heisenbergstra{\ss}e 1, 70569 Stuttgart, Germany}
\author{Santiago Blanco-Canosa}
\affiliation{Donostia International Physics Center, DIPC, 20018 Donostia-San Sebastian, Basque Country, Spain}
\affiliation{IKERBASQUE, Basque Foundation for Science, 48013 Bilbao, Basque Country, Spain}
\author{Georg Christiani}
\affiliation{Max Planck Institute for Solid State Research, Heisenbergstra{\ss}e 1, 70569 Stuttgart, Germany}
\author{Gennady Logvenov}
\affiliation{Max Planck Institute for Solid State Research, Heisenbergstra{\ss}e 1, 70569 Stuttgart, Germany}
\author{Bernhard Keimer}
\affiliation{Max Planck Institute for Solid State Research, Heisenbergstra{\ss}e 1, 70569 Stuttgart, Germany}
\author{Matteo Minola}
\email[Email address: ]{m.minola@fkf.mpg.de}
\affiliation{Max Planck Institute for Solid State Research, Heisenbergstra{\ss}e 1, 70569 Stuttgart, Germany}

\begin{abstract}
We used resonant elastic x-ray scattering at the \ce{Cu} $L_3$ and \ce{Dy} $M_5$ edges to
investigate charge order in thin films of underdoped \ce{DyBa2Cu3O_{6+x}} (DyBCO)
epitaxially grown on \ce{NdGaO3} (110) substrates. The films show an orthorhombic crystal
structure with short-range ortho-II oxygen order in the charge-reservoir layers. At the
Dy $M_5$ edge we observe diffraction peaks with the same planar wavevectors as those of
the two-dimensional charge density wave in the \ce{CuO2} planes and of the ortho-II
oxygen order, indicating the formation of induced ordered states on the rare-earth
sublattice. The intensity of the resonant diffraction peaks exhibits a non-monotonic
dependence on an external magnetic field. Model calculations on the modulation of the
crystalline electric field at the Dy sites by charge and oxygen order capture the salient
features of the magnetic field, temperature, and photon energy dependence of the
scattering intensity.
 \end{abstract}
\pacs{not.a.number}
\date{\today}

\maketitle

\section{Introduction}
\label{sec:introduction}

Strongly correlated compounds are often characterized by a great variety of phases which
are frequently in competition with each other. The high-temperature superconducting
cuprates represent a particularly interesting case given the richness of their phase
diagram, with antiferromagnetism, superconductivity, electronic nematicity and charge
order in close proximity of one another\cite{Keimer2015}. In this work, we investigated
the charge ordered state in \ce{DyBa2Cu3O_{6+x}} (DyBCO) thin films using resonant x-ray
scattering (RXS). The ``123'' family of cuprates \emph{R}\ce{Ba2Cu3O_{6+x}} ($R$BCO),
where $R$ is a rare-earth, is best known for the compound YBCO, which has been
extensively investigated in recent years, owing to its high superconducting critical
temperature ($T_\textrm{c}$) and very low chemical disorder. In addition to the
superconducting \ce{CuO2} planes, the crystal structure of these materials comprises a
charge-reservoir layer with \ce{CuO} chains. Hole doping of the \ce{CuO2} plane can be
achieved with relatively little associated disorder by introducing additional oxygen ions
into the charge-reservoir layer. In underdoped YBCO, the \ce{CuO} chains run along the
$b$ direction and organize in regular sequences of filled and empty sites along the $a$
direction, resulting in a uniaxial superstructure (termed ortho-ordering) whose spatial
period depends on the doping\cite{Zimmermann2003}.

RXS at the Cu $L_3$ edge has proven to be an efficient method of investigating charge
density wave (CDW) order because it is sensitive enough to detect short-range
correlations and CDW fluctuations. In underdoped YBCO, charge order diffraction peaks
have been observed along both planar bond directions\cite{Ghiringhelli2012}, and have
been interpreted in terms of perpendicular domains of uniaxial CDW
correlations\cite{Comin2015,Kim3DCDW}. At zero magnetic field, and in the absence of
external pressures, this charge order is confined to the \ce{CuO2} planes, with little
correlation along the $c$ direction\cite{Chang2012}. The temperature dependence of the
CDW diffraction peaks in YBCO shows a maximum in the scattering intensity for $T =
T_\textrm{c}$. At the same temperature, the correlation length $\xi$ reaches a maximum of
$\sim 20$ lattice parameters\cite{BlancoCanosa2013}. This behavior implies a strong
competition between charge order and superconductivity in YBCO. The incident photon
energy dependence of the CDW peaks also clearly shows that only the Cu ions in the
\ce{CuO2} planes (with a valence state close to Cu$^{2+}$) take part in the charge
ordering phenomena, while the Cu$^{1+}$ atoms in the charge-reservoir layers are not
involved. The converse is true for the ortho-order peaks. The Y$^{3+}$ ions (with a
full-shell electron configuration) do not participate in the charge ordering.

In addition to the quasi-two-dimensional (quasi-2D) CDW and superconductivity, several
recent studies have demonstrated that a long-range three-dimensional (3D) charge ordered
state can be stabilized in underdoped YBCO by the application of  various external
fields\cite{Frano2020}. This 3D CDW manifests itself as a sharp diffraction peak at
integer values of the out-of-plane Miller index $L$, and appears to coexist with the 2D
CDW, whose diffraction signatures remain intact across the 3D CDW transition. First, it
was found that by applying high external magnetic fields along the $c$
axis\cite{Gerber2015,Chang2016} ($\geq \SI{15}{\tesla}$), the 3D CDW forms only along the
$b$ axis, in concomitance with the suppression of superconductivity. More recently, it
has been shown that a similar 3D charge order can also be induced in YBCO crystals by
uniaxial strain along the $a$ axis\cite{Kim2018,Kim3DCDW} and by epitaxial growth of thin
films on cubic \ce{SrTiO3} (STO) substrates\cite{Bluschke2018}. In all cases, the 3D CDW
shares the same in-plane wave vector with its 2D counterpart. In contrast to the case of
bulk YBCO, epitaxially induced 3D CDW correlations are extraordinarily robust,
demonstrating little or no suppression upon entering the superconducting phase and
surviving up to and above room temperature. These findings point towards a structural
mechanism stabilizing the 3D CDW in YBCO thin films. An independent study on the related
``123'' cuprate material \ce{(Ca_{x}La_{1-x})(Ba_{1.75-x}La_{0.25+x})Cu3O_{y}} (CLBLCO)
showed that structural degrees of freedom  couple strongly to the 2D CDW as isovalent
chemical substitution on the $A$ and $B$ sites was used to redistribute chemical pressure
within the unit cell\cite{Bluschke2019}.

$R$BCO compounds with  $R \neq$ Y differ from YBCO in two respects. First, the different
sizes of the $R^{3+}$ ions modify the interatomic distances and associated electronic
hopping parameters, resulting in slight variations of the electronic properties,
including the superconducting $T_\textrm{c}$. Second, the partially filled $f$-shells
generate magnetic moments that are, however, only weakly coupled to the conduction
electrons in the \ce{CuO2} planes. (Note that the $R$ = Pr compound is anomalous because
of deviations from the 3+ valence state and/or hybridization of Pr orbitals with
electronic states in the \ce{CuO2} planes.)

To investigate the impact of these effects on CDW order, we have performed RXS
experiments on DyBCO thin films synthesized by molecular beam epitaxy\cite{Putzky2020}.
The electronic properties of the Dy$3^+$ ions (electron configuration $4f^9$) in DyBCO
have been studied in bulk DyBCO crystals by paramagnetic resonance and neutron
scattering\cite{Likodimos2001,Allensnpach1989}. In the $D_{2h}$ point group symmetry, the
ground state is composed almost exclusively of the $M_J = \pm \nicefrac{11}{2}$ component
of the $J = \nicefrac{15}{2}$ multiplet. The magnetic moments generated in this way
interact weakly and order antiferromagnetically at \SI{1}{\kelvin}\cite{Goldman1987}.
Since the presence of ortho-order is known to modify the crystal field environment
surrounding the Dy ions\cite{Allenspach1989_2}, we used Dy $M$-edge RXS to investigate
possible imprints of CDW and oxygen order on the resonant diffraction cross section.

\begin{figure}
\centering
\includegraphics[width=\columnwidth]{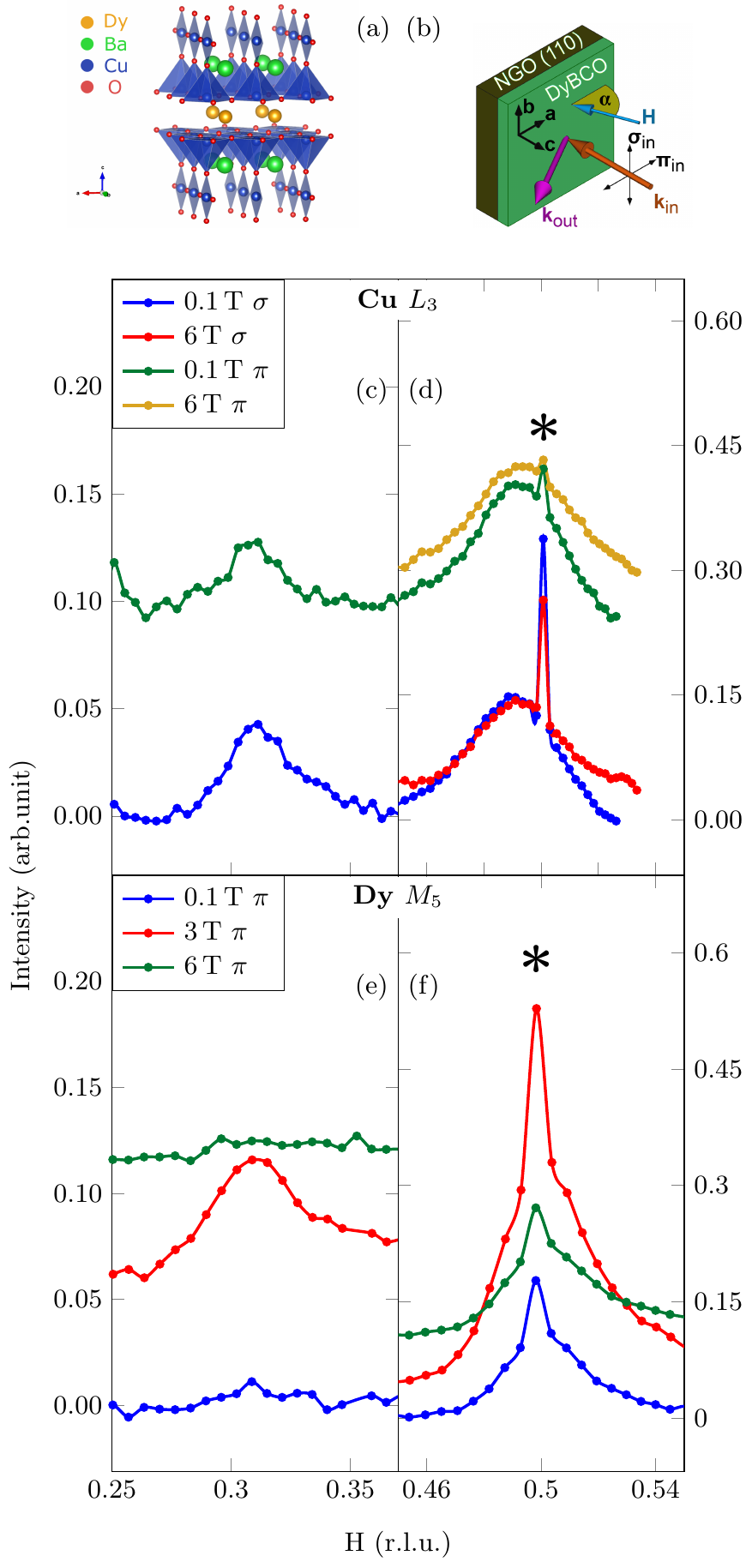}
\caption{(a): DyBCO crystal structure. (b): Sketch of the experimental scattering geometry.
(c-f): Rocking scans (converted to reciprocal-space coordinate H) for DyBCO at the Cu $L_3$ (c-d) and Dy $M_5$ (e-f) edges at different magnetic field values. The curves
have been shifted vertically for clarity. The spikes marked by asterisks in panels (d,f) at
exactly $q = (0.5, 0.0)$ arise from tails of a very intense nearby substrate peak at (0.5, 0.0, 2.0) (pseudocubic notation) and should therefore be ignored.
 } \label{fig:longscans}
\end{figure}

In our study of DyBCO thin films we have first used Cu $L_3$-edge RXS to detect a 2D CDW
that resonates with the Cu$^{2+}$ ions of the \ce{CuO2} planes, analogous to prior work
on YBCO\cite{BlancoCanosa2014}. Surprisingly, we also observed an ortho-II \ce{CuO} chain
order in the charge reservoir layer, a phenomenon which has not previously been observed
in cuprate thin films. When tuning the incident photon energy to the Dy $M_5$ resonance,
diffraction peaks are observed at the wave vectors associated with both the CDW and the
\ce{CuO} chain order. The temperature, field and photon energy dependence of these
reflections are consistent with those of an induced ordering phenomenon. We also report
the absence of 3D CDW correlations in our films and discuss the possible role of
structural degrees of freedom in suppressing the 3D ordered state.

\begin{figure}
\centering
\includegraphics[width=\columnwidth]{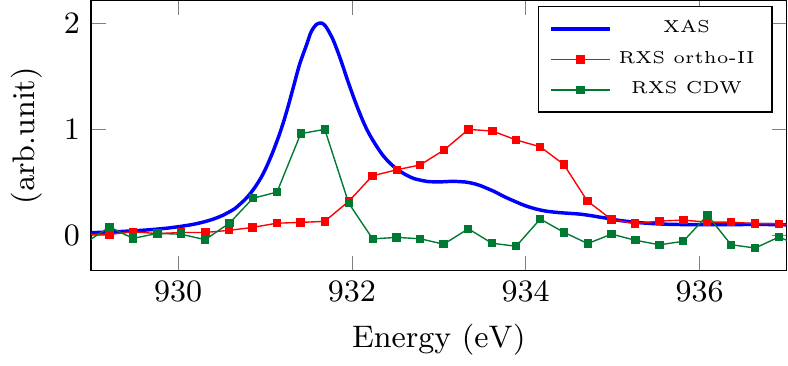}
\caption{Cu $L_3$ incident photon energy dependence of the RXS reflections at $q = (0.5, 0.0)$ and (0.315, 0.0).
The spectra were taken at \SI{5}{\kelvin} with $\sigma$ polarization of the x-rays and no external magnetic field.
 } \label{fig:escansCu}
\end{figure}

\section{Experimental details}
\label{sec:experimental_details}

The thin films of DyBCO were grown on orthorhombic \ce{NdGaO3} (NGO) (110) substrates
using layer-by-layer molecular beam epitaxy, which produces films with the $c$ axis
perpendicular to the surface. The crystalline quality was characterized by x-ray
diffraction and scanning transmission electron microscopy, while $T_\textrm{c}$ was
obtained by mutual inductance measurements. We measured two films, both
\SI{20}{\unitcell} thick. Both samples were reduced after growth resulting in
$T_\textrm{c} \sim \SI{55 \pm 5}{\kelvin}$, which corresponds to approximately $x=0.5$ in
bulk crystals. However, repeated characterization of the superconducting critical
temperature revealed a further decrease in the oxygen content over a time scale of days
and weeks. Nevertheless, the presence of both ortho-II and CDW ordering reflections in
our RXS experiments verify that the doping level of our samples remained in the
underdoped regime during the measurements presented here. We observed no sign of changes
due to beam damage. The films show orthorhombic twin domains as previously
reported\cite{Steinborn1994}, thereby making the two in-plane directions equivalent in
our measurements, which average over a $\sim \SI{200}{\micro\metre}$ beam spot. Details
of the growth and characterization are given in Ref.\,\onlinecite{Putzky2020}.

The x-ray measurements were performed at the UE46 PGM-1 beamline of the Helmholtz-Zentrum
Berlin at BESSY-II. The experimental station comprises two different ultra-high-vacuum
chambers, one of which allows for the application of an external magnetic field up to
\SI{6}{\tesla} using an in-vacuum rotatable superconducting coil. The rotation of this
coil, and therefore the field direction, is limited to a set of angular ranges for which
the coil does not interfere with the x-ray beam bath. For the RXS measurements of the CDW
at the Dy $M_5$ edge, the applied field was kept at $\sim \SI{45}{\degree}$ from the
in-plane direction in the $a-c$ plane (angle $\alpha$ in Fig.\,\ref{fig:longscans} (b)),
which is the lowest angle allowed by the geometry. For the ortho-order measurement at the
same edge, the field was directed at $\alpha \sim \SI{30}{\degree}$.
The liquid helium cryostat produces a sample environment with temperatures as low as
\SI{5}{\kelvin}. The incoming light polarization was either parallel ($\pi$) or
perpendicular ($\sigma$) to the horizontal scattering plane.


\begin{figure}
\centering
\includegraphics[width=\columnwidth]{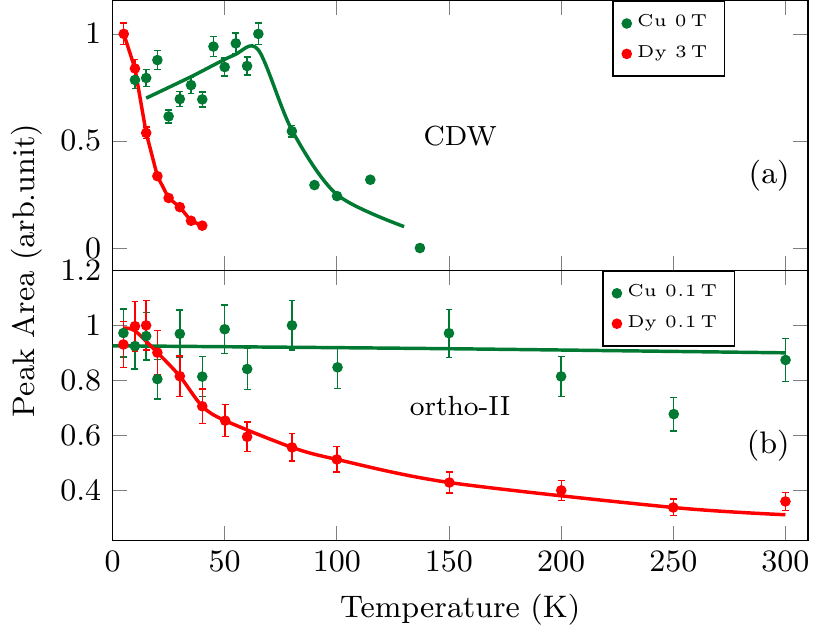}
\caption{Temperature dependence of the charge order (a) and ortho-order (b) peaks. The maximum intensity of each data set is normalized to 1. Lines are guides to the eye.
 } \label{fig:tempdep}
\end{figure}

\begin{figure}
\centering
\includegraphics[width=\columnwidth]{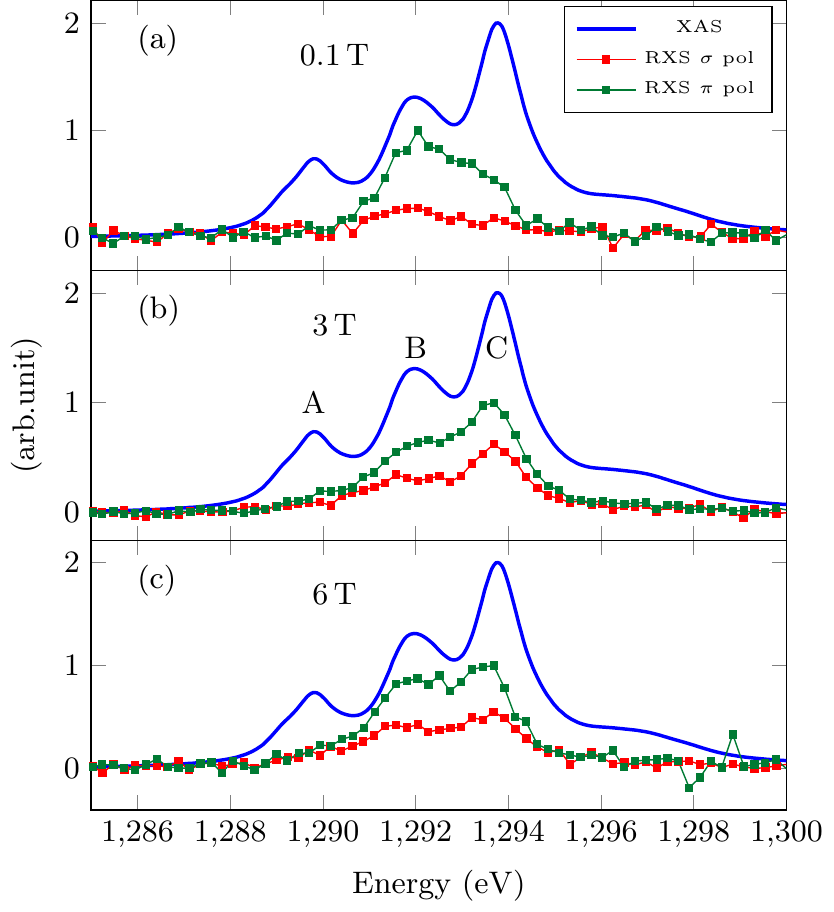}
\caption{Dy $M_5$ incident photon energy dependence of the RXS reflection at (0.5, 0.0) for increasing magnetic field.
The intensities of the XAS and the RXS spectra are normalized to 2 and 1 (for $\pi$ polarization), respectively. All data was taken at \SI{5}{\kelvin}.
 } \label{fig:escansDyOrtho}
\end{figure}

RXS is a photon-in-photon-out technique where the scattering signal is greatly enhanced
by tuning the incoming energy to a particular absorption edge of an ion in the compound.
When the incident photon energy is tuned to an absorption edge, the photon energy
satisfies a resonance condition which strongly couples a core electronic state to the
valence states during the scattering process. RXS is therefore sensitive to variations in
the charge and/or magnetic degrees of freedom in the valence electrons of the scattering
ions. In $R$BCO compounds, there are two inequivalent Cu ions in the structure, one in
the \ce{CuO2} planes (approximately Cu$^{2+}$) and one in the chains (approximately
Cu$^{1+}$), with the exact valence states depending on the doping level. These distinct
sites give rise to specific features in the absorption spectra at the Cu $L_3$ resonance
($\sim \SI{931}{\electronvolt}$)\cite{Hawthorn2011}. Accordingly, measuring the
diffraction peak's incident photon energy dependence in the vicinity of an absorption
edge allows us to determine which electronic states, and thus which ions, are involved in
the diffracting superstructure. In addition, at the Dy $M_5$ resonance ($
\SI{1292}{\electronvolt}$) it is also possible to determine if a diffraction peak
originates from orbital or magnetic ordering by considering the detailed RXS
spectroscopic lineshape\cite{Schierle2010}. The x-ray absorption spectra (XAS) consists
of three peaks at $\sim$ 1290, 1292 and \SI{1294}{\electronvolt}, which we label A, B and
C (see Fig.\,\ref{fig:escansDyOrtho}). Peaks A and C correspond to $\Delta J = \pm 1$
transitions, while peak B corresponds to $\Delta J=0$ transitions. Each $J$-level is also
split into $M_J$ levels by the crystal field or by applied magnetic fields, with the
additional selection rule $\Delta M_J = 0,\pm 1$ applying to electric-dipole-allowed
transitions. In Dy ions each $\Delta J$ transition is dominated by a single $\Delta M_J$
value and therefore the latter selection rule approximates that of $\Delta
J$\cite{Ott2006}. It follows that diffraction peaks which originate from magnetic
ordering of the $M_J$ moments will resonate almost exclusively at peaks A and C, while a
structural diffraction peak or a peak originating from charge or orbital ordering will
result in a resonance at peak B\cite{Schierle2010}.

\begin{figure}
\centering
\includegraphics[width=\columnwidth]{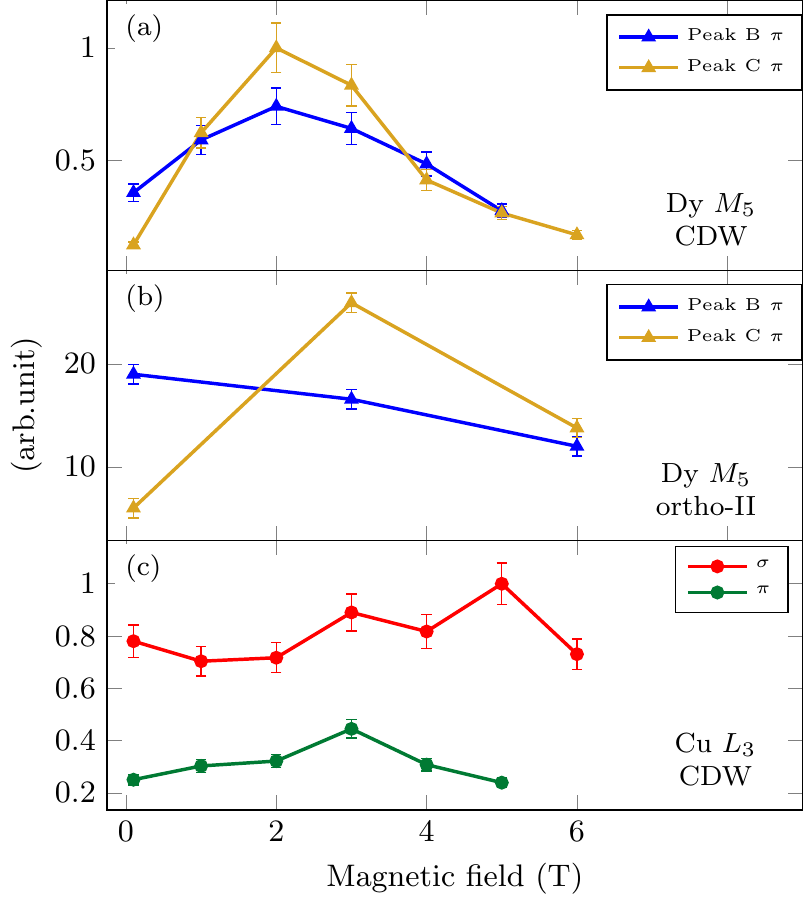}
\caption{Magnetic field dependence at the Dy $M_5$ edge of (a) the CDW peak and (b) the ortho-II peak; (c) the CDW peak at the Cu $L_3$ edge.
Peak C of the XAS (energy \SI{1294}{\electronvolt}) is of magnetic origin, while peak B (energy \SI{1292}{\electronvolt})
is of charge, orbital or structural origin\cite{Schierle2010}. The magnetic field was applied at an angle $\alpha \sim \SI{45}{\degree}$ for the Dy CDW peak in panel (a), at
$\alpha \sim \SI{30}{\degree}$ for the ortho-II peak in panel (b) and at
$\alpha \sim \SI{45}{\degree}$ for the Cu CDW peak in panel (c). See Fig.\,\ref{fig:longscans} (b) for a sketch of the experimental scattering geometry.
All data was taken at \SI{5}{\kelvin}.
 } \label{fig:fielddep}
\end{figure}

\section{Results}
\label{sec:results}

Fig.\,\ref{fig:longscans} shows rocking scans projected onto the [10] direction of the
DyBCO thin films at both the Cu $L_3$ and Dy $M_5$ edges for increasing external magnetic
fields. The scans at the Cu edge show a CDW peak at $q \sim (0.315, 0.0)$, as expected,
and, remarkably, also a broad ortho-II peak close to $q = (0.5, 0.0)$. Throughout this
manuscript we will refer to positions in the 2D reciprocal space (H, K) in units of
$(2\pi/a,2\pi/b)$ (reciprocal lattice units, \si{\rlu}) with $a$ and $b$ the in-plane
lattice parameters. The incident photon energy dependence of the peak intensities is
shown in Fig.\,\ref{fig:escansCu}. Each peak exhibits a characteristic resonance profile.
The CDW, associated with the Cu$^{2+}$ ions of the \ce{CuO2} planes, is peaked near
\SI{931.7}{\electronvolt}. In contrast, the ortho-II \ce{CuO} chain order reflection is
associated with copper ions in the charge reservoir layer and peaks at higher energies
near \SI{933.2}{\electronvolt}. Contrary to similar YBCO thin films grown on
STO\cite{Bluschke2018}, no signature of a 3D CDW was found. While it has been shown that
$R$BCO compounds can grow in the orthorhombic structure in thin film
form\cite{Steinborn1994,Putzky2020}, the presence of ortho-order \ce{CuO} chains is
remarkable since these have not been reported in $R$BCO thin films to date.

The most interesting result, however, is found when the incident photon energy is tuned
to the Dy $M_5$ edge. Both the CDW and the ortho-II order peaks are detected on the Dy
sublattice, with maximum intensity around \SIrange{2}{3}{\tesla}, indicating that the
modulation in the Cu atoms is transferred to the Dy ions. The correlation lengths at
\SI{5}{\kelvin} are $\sim \SI{27}{\angstrom}$ and $\sim \SI{32}{\angstrom}$ for the
induced ortho and CDW orders, respectively. In various transition metal
oxides\cite{Prokhnenko2007,Donnerer2019} and other multinary rare-earth containing
systems, such as the rare-earth tritellurides\cite{Lee2012} or synthetic multilayer
systems\cite{Bluschke2017}, the formation of induced order on the rare-earth sublattice
has already been observed. Likewise, in DyBCO, the temperature-dependent resonant
scattering intensity at the Dy edge decays quickly with temperature, reflecting the
temperature-dependent susceptibility of the Dy sublattice (Fig.\,\ref{fig:tempdep} (a)).
This is in contrast to the behavior of the two corresponding reflections measured at the
Cu $L_3$ edge, where the CDW peak decreases rapidly in intensity only above the
superconducting transition\cite{BlancoCanosa2014} and the ortho-order peak is almost
constant up to room temperature (Fig.\,\ref{fig:tempdep} (b)), in agreement with prior
work on bulk single crystals\cite{BlancoCanosa2014}.

\begin{figure}
\centering
\includegraphics[width=\columnwidth]{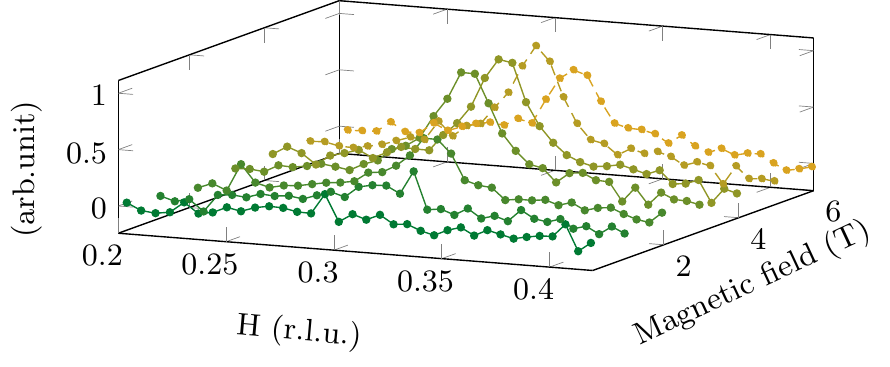}
\caption{Magnetic field dependence of the charge order peak at the Dy $M_5$ edge, taken at \SI{5}{\kelvin}.
The data is normalized to 1.
 } \label{fig:waterfalls}
\end{figure}

\begin{figure}
\centering
\includegraphics[width=\columnwidth]{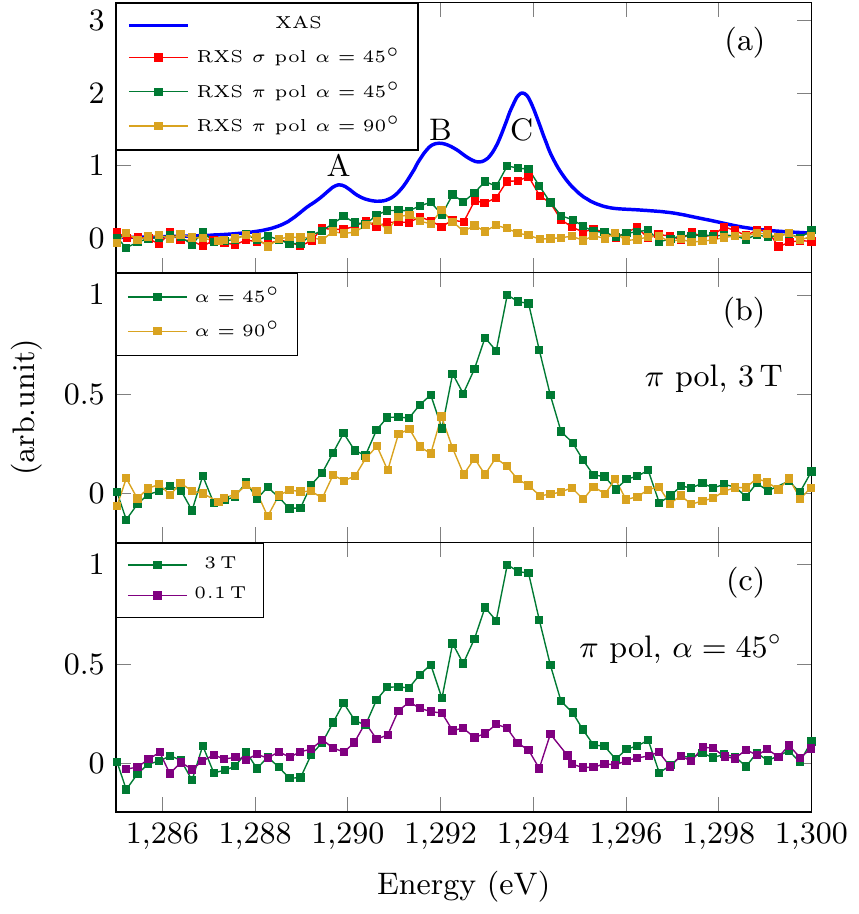}
\caption{(a): Dy $M_5$ incident photon energy dependence of the RXS reflection at the CDW $q = (0.315, 0.0)$. All data in this panel was taken at \SI{3}{\tesla} but it
is representative for field values $1-\SI{6}{\tesla}$: the intensity changes with field but the spectral shape does not.
The intensities of the XAS and the RXS spectra have each been scaled in order to facilitate a comparison of their lineshapes.
(b): Comparison between the RXS signal for different $\alpha$ values.
(c): Comparison between the RXS signal at high and low magnetic field.
All data was taken at \SI{5}{\kelvin}.
 } \label{fig:escansDychargeorder}
\end{figure}

\begin{figure*}
\centering
\includegraphics[width=\textwidth]{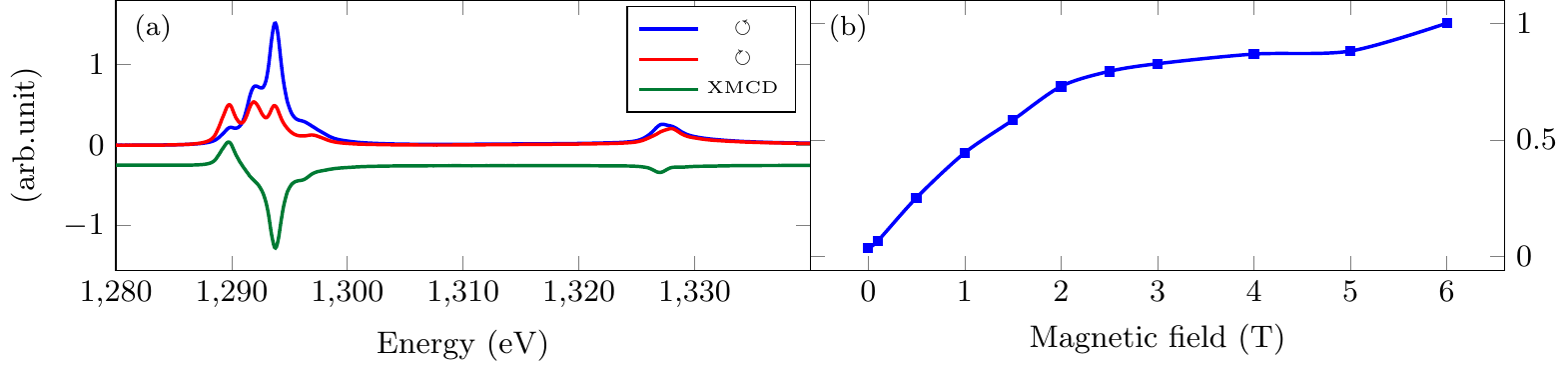}
\caption{XAS and XMCD spectra at \SI{6}{\tesla} (a) and field dependence of the
integrated dichroic signal (b). The XMCD curve in (a) has been shifted downwards for
clarity. The data was taken at \SI{5}{\kelvin}.
 } \label{fig:xmcd}
\end{figure*}

\begin{figure}
\centering
\includegraphics[width=\columnwidth]{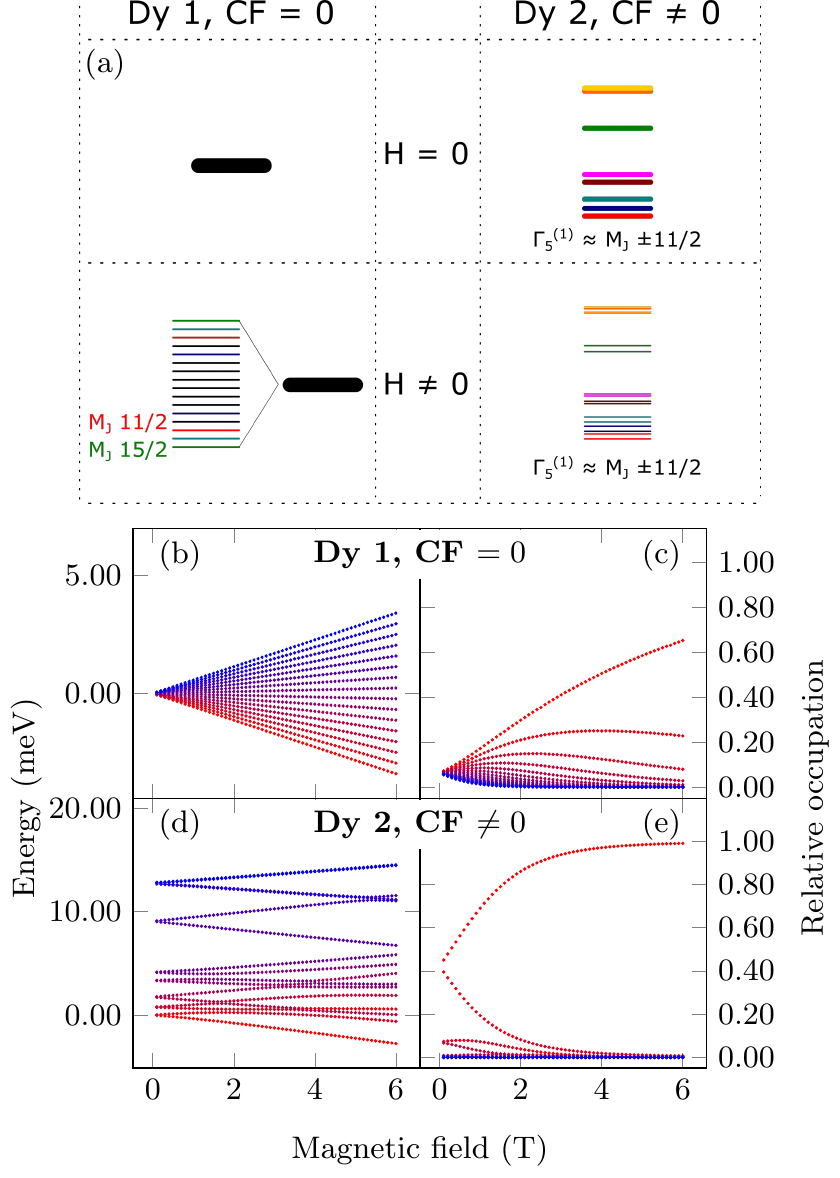}
\caption{(a): Sketch of the Dy electronic levels for the two different sites (with and without crystal field (CF) and external magnetic field). The crystal field
and the magnetic field tend to stabilize different electronic levels as ground states.
It is worth noticing that the $\Gamma$ multiplets only roughly correspond to the $M_J$ states\cite{Allensnpach1989}.
(b): Energy levels for a Dy ion without crystal field (Dy 1) as a function of the applied magnetic field at $\alpha = \SI{45}{\degree}$.
(c): Boltzmann occupation of the levels in (b) at \SI{5}{\kelvin}.
(d): Energy levels for a Dy ion with crystal field (Dy 2) as a function of the applied magnetic field at $\alpha = \SI{45}{\degree}$.
(e): Boltzmann occupation of the levels in (d) at \SI{5}{\kelvin}.
} \label{fig:levels_sketch}
\end{figure}

\begin{figure}
\centering
\includegraphics[width=\columnwidth]{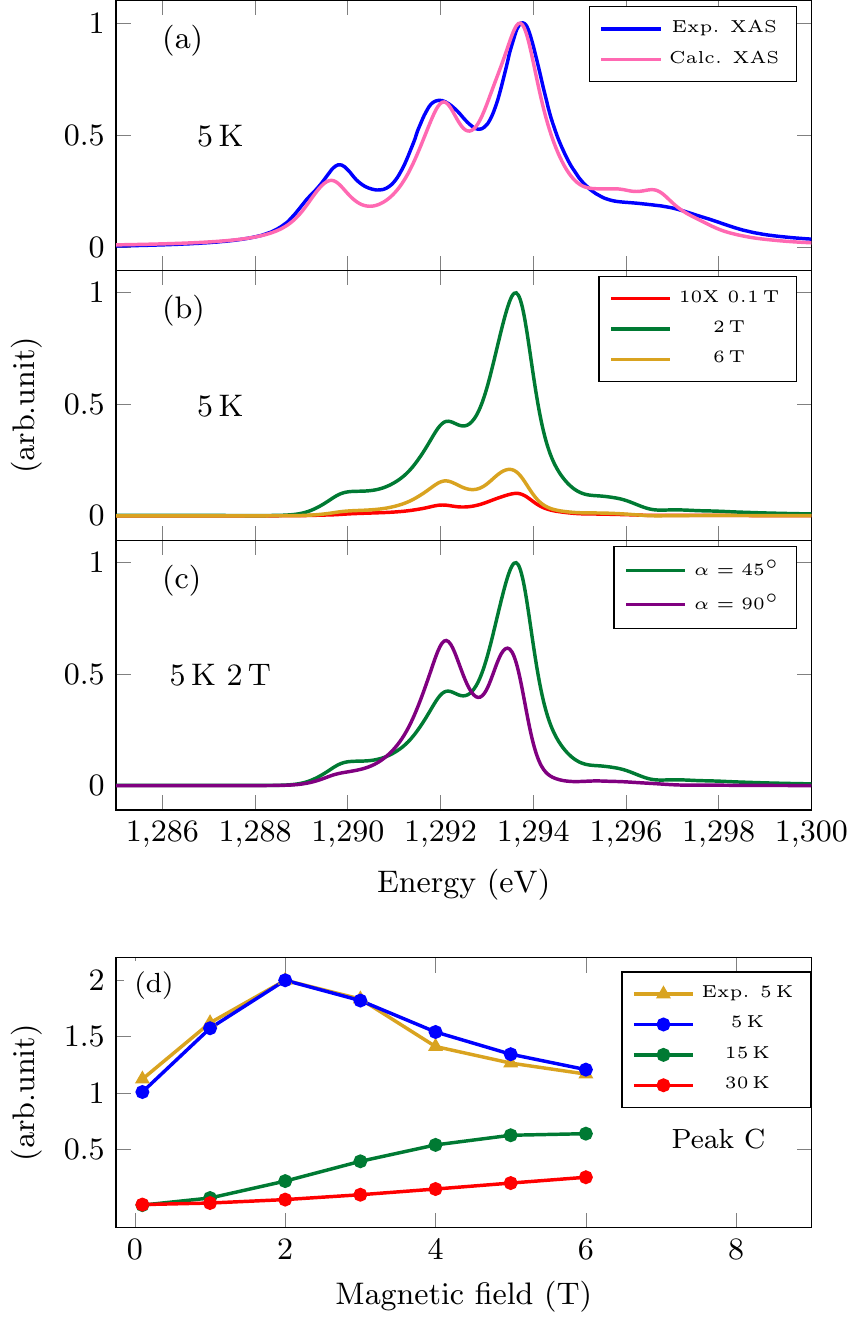}
\caption{(a): Comparison between the experimental and calculated XAS spectra. (b): Calculations for the RXS spectral shape at \SI{5}{\kelvin} for three values of magnetic field with
$\alpha = \SI{45}{\degree}$. In (c) we compare the signal for $\alpha = \SI{45}{\degree}$ and $\alpha = \SI{90}{\degree}$ (out-of-plane field).
(d): Field dependence of the RXS intensity at peak C for increasing temperature. We show the \SI{5}{\kelvin} curve together with the experimental data for the same
temperature (Fig.\,\ref{fig:fielddep} (a)). Both curves have been shifted vertically for clarity.
See text for details of the calculations.
 } \label{fig:sims}
\end{figure}


The photon energy dependence of the ortho-II peak at the Dy $M_5$ edge is shown in
Fig.\,\ref{fig:escansDyOrtho} for different magnetic fields. Our data clearly indicates
that this induced order has an important contribution of magnetic origin (peak C) but
there is also possibly a contribution from orbital or charge ordering (peak B).
Fig.\,\ref{fig:fielddep} (b) demonstrates that the resonance at energies corresponding to
peak C of the XAS has a non-monotonic field dependence, with its maximum occurring near
\SI{3}{\tesla}.

We now turn our attention to the charge order peak at $q = (0.315, 0.0)$ measured at the
Dy $M_5$ edge (see Fig.\,\ref{fig:waterfalls} for the field dependence plot). At low
fields (\SI{0.1}{\tesla}), the energy dependence of the reflection, shown in
Fig.\,\ref{fig:escansDychargeorder} (c), demonstrates essentially no scattering at photon
energies corresponding to peak C. However, it should be noted that the overall signal is
weakest at low fields, possibly hampering our ability to accurately determine the
resonant line shape. At higher fields, the contribution of peak C becomes dominant,
indicating enhanced magnetic scattering. The total intensity peaks around
\SIrange{2}{3}{\tesla} and is suppressed again at higher fields (see
Fig.\,\ref{fig:fielddep} (a)).
 A similar dependence of the RXS signal on the field strength has been
observed by \citet{Donnerer2019} in \ce{Tb2Ir2O7}.
Finally, we observed that modifying the field direction from $\alpha \sim
\SI{45}{\degree}$ towards a fully out-of-plane configuration, \emph{i.e.} along the $c$
axis, severely reduces the scattering signal (Fig.\,\ref{fig:escansDychargeorder} (b)).

\section{Discussion}
\label{sec:discussion}

Our RXS study of DyBCO films shows a 2D CDW in the \ce{CuO2} planes without any evidence
for long-range correlations along the $c$ direction (\emph{i.e.} no 3D CDW). These
findings are analogous to bulk YBCO but differ from a related report on YBCO films grown
on \ce{SrTiO3}, for which a 3D CDW was observed in zero field\cite{Bluschke2018}. In
Ref.\,\onlinecite{Bluschke2018} it is argued that the suppression of orthorhombic crystal
symmetry, via the epitaxial relationship to the cubic substrate, may promote 3D CDW
correlations in the YBCO thin films. According to this picture the bulk-like \ce{CuO}
chain order and associated orthorhombic crystal symmetry of our DyBCO films may be
responsible for inhibiting the formation of a 3D CDW. However, other degrees of freedom
are manipulated upon substitution of Y for Dy. Not least among these is the presence of a
large $4f$ moment on the Dy site, which we have shown is coupled to the CDW.
Additionally, small variations in bond angles and lengths associated with the smaller
size of the Dy ion may have important implications for the charge ordered state, similar
to the structural modifications achieved in CLBLCO upon isovalent chemical
substitution\cite{Bluschke2019}. Further work is required to elucidate the full impact of
epitaxial growth and structural degrees of freedom on the charge ordered state in
cuprates.

Our magnetic field-dependent measurements clearly suggest that the magnetic moment of the
Dy ions must play a crucial role in coupling the rare-earth sublattice to the electronic
and structural degrees of freedom in \ce{CuO2} planes and in the charge-reservoir layer.
\citet{Lee2012} described the formation of an induced charge order on the rare-earth
atoms in tritellurides as a periodic modulation of the crystal field acting on the
different $M_J$ levels of the spin-orbit coupled electronic ground state. We performed
calculations akin to the ones in \citet{Lee2012} using the Quanty
code\cite{Haverkort2012}. The scattering tensor for a Dy$^{3+}$ ion in a $D_{2h}$ crystal
field (according to Ref.\,\onlinecite{Allensnpach1989}) is calculated for different
magnetic field values. Slater integrals and spin-orbit coupling parameters are calculated
using Cowan's code\cite{Cowan1981}. Boltzmann averaging is performed to obtain the
thermal occupation of the lowest energy states.

As the CDW is incommensurate and involves motion of multiple ions in the unit cell,
accurate modeling of the corresponding crystalline electric field at the Dy site is a
formidable task. To capture the essence of the mechanism underlying the imprint of CDW
order on the Dy site, we adopt a simple model with only two Dy sites, which can be
thought of as those experiencing the maximum and minimum CDW-induced modulation. For
simplicity, the crystal field acting on one of these ions is assumed to be identical to
the one described by parameters extracted from neutron scattering experiments, apart from
a rescaling factor of 0.25\cite{Allensnpach1989}, and the other one is assumed to be
isotropic. The modulation of the scattering tensor between these two sites is used to
represent the periodic modulation of the crystal field parameters associated with the
charge density wave (or ortho-order), and the resonant scattering profile is obtained by
taking the difference of the scattering tensors at the two sites.

At one site the Dy crystal field parameters stabilize the $M_J = \pm\nicefrac{11}{2}$
ground state in zero field, whereas on the other site all $M_J$ states are degenerate. An
external magnetic field then splits all Kramer's doublets via the Zeeman effect resulting
in an $M_J = \nicefrac{15}{2}$ ground state for sufficiently high fields, \emph{i.e.} the
state with the largest magnetic moment component along the field direction. For
intermediate fields, both sites exhibit a net moment in the direction of the field, but
the magnitude of the moments is modulated by the distinct crystal field environments at
the two sites. Fig.\,\ref{fig:levels_sketch} (a) shows a sketch of the resulting
crystal-field eigenstates with and without magnetic field for the Dy ions at the two
sites. The mechanism behind the peculiar non-monotonic field dependence of the induced
order is then captured by panels (b-e), which show the calculated field dependence of the
states and the corresponding Boltzmann occupation at \SI{5}{\kelvin} for zero (b-c) and
non-zero crystal field (d-e). The RXS intensity results then from the contrast arising
from the small changes in the level occupations and the XAS final state of the two Dy
sites. Notice in particular in Fig.\,\ref{fig:levels_sketch} (d) the crossing of the
low-lying energy levels around \SIrange{2}{3}{\tesla}, which, in concert with the
evolving level occupations, is likely responsible for the maximum at the same field value
of the RXS signal measured on peak C (Fig.\,\ref{fig:fielddep} (a)). In the real material
we believe that the magnetic component of the resonant scattering cross section arises
from field-aligned Dy moments whose magnitudes are modulated at the CDW (or ortho-order)
wavevector by the CDW (or ortho-order) induced modulations of the crystal field.

The results of our calculations are shown in Fig.\,\ref{fig:sims}. The spectral shape of
the resonance at the different field and temperature values qualitatively reproduces the
experimental data for both induced CDW and ortho-order, mimicking well field as well as
temperature dependences, with a reasonable, albeit rough, quantitative agreement. In
particular, the striking observation of a non-monotonic field dependence of the RXS
intensity is reproduced at low temperatures corresponding to the experimental conditions.
Fig.\,\ref{fig:sims} (b) ($T = \SI{5}{\kelvin}$) clearly shows that the RXS signal at
\SI{2}{\tesla} is significantly higher compared to \SI{0.1}{\tesla} and \SI{6}{\tesla}.
The complete field dependence shown in Fig.\,\ref{fig:sims} (d) for three temperatures is
in very good agreement with the experimental data. We interpret this non-monotonic
behavior in terms of two regimes. In the low-field regime the amplitude of the
CDW-induced (or ortho-order induced) modulation of the local Dy moments reflects the
magnitude of the Dy sublattice magnetization and thus grows with increasing field
strength. For higher field strengths the Dy magnetization is essentially saturated.
Further increases in the field only serve to bring the states preferred by the Zeeman
splitting so far down in energy that the RXS intensity at the CDW (or ortho-order) wave
vector is suppressed. This picture is supported by XMCD measurements performed at the Dy
$M_5$ edge (Fig. 8), which demonstrate paramagnetic behavior with saturation at $\sim
\SIrange{2}{3}{\tesla}$. Our model further captures the dramatic reduction of the RXS
intensity for photon energies corresponding to peak C in
Fig.\,\ref{fig:escansDychargeorder} (a), when the magnetic field is applied along the
out-of-plane direction ($\alpha = \SI{90}{\degree}$) compared to $\alpha =
\SI{45}{\degree}$. This is shown in Fig.\,\ref{fig:sims} (c). A complete quantitative
agreement with the data would involve the optimization of a great number of model
parameters, and would likely be complicated by the presence of disorder in the system.
Accordingly, a comprehensive model exceeds the scope of this study and may be the subject
of future work.

\section{Conclusions}
\label{sec:conclusions}

We have reported RXS measurements of underdoped thin films of the high-temperature
superconductor \ce{DyBa2Cu3O_{6+x}} grown on \ce{NdGaO3} (110) substrates. The films were
found to host both ortho-II oxygen order as well as a 2D CDW, akin to those observed in
bulk underdoped YBCO but with shorter correlation lengths. Unlike thin films of YBCO,
which were recently reported to host 3D CDW correlations at zero
field\cite{Bluschke2018}, the CDW observed in our films was found to be only weakly
correlated along the $c$ direction. Our central result is the additional observation of
both the ortho-II order and the CDW reflections for incident photon energies tuned to the
Dy $M_5$ resonance. The temperature dependence of these diffraction peaks reveals the
induced nature of these ordering tendencies, while both the photon energy and magnetic
field dependences indicate a magnetic contribution to the resonant scattering cross
section. The magnetic moments of the Dy ions in the ``123'' cuprate structure can thus be
used as a sensitive probe of both structural ordering tendencies, as well as the
electronic correlations in the \ce{CuO2} planes. In the related system
\ce{PrBa2Cu3O_{6+x}}, a complex interaction between the Pr $4f$ moments and the \ce{CuO2}
planes has previously been reported\cite{Hill1998}. Future studies of the charge ordered
state in this non-superconducting system may reveal further insights into the role of the
4$f$ moments in coupling to both the CDW and superconductivity.

\section*{Acknowledgments}
We acknowledge the financial support from the Deutsche Forschungsgemeinschaft (DFG,
German Research Foundation), Projekt No.107745057 TRR 80, from the European Union's
Horizon 2020 research and innovation programme under Grant Agreement No. 823717- ESTEEM3
and from the Alexander von Humboldt Foundation.

\bibliography{cuprates}
\end{document}